\documentclass[tightenlines,superscriptaddress,amsmath,amssymb,prl,12pt]{revtex4-1}
\usepackage{url}
\usepackage{graphicx}
\usepackage{dcolumn}
\usepackage{bm}
\usepackage{mathptmx}
\usepackage{xcolor}
\usepackage{ulem}

\newcommand{\msun}{{\rm M}_\odot}

\newcommand\lsim{\mathrel{\rlap{\lower4pt\hbox{\hskip1pt$\sim$}}
        \raise1pt\hbox{$<$}}}
\newcommand\gsim{\mathrel{\rlap{\lower4pt\hbox{\hskip1pt$\sim$}}
        \raise1pt\hbox{$>$}}}

\begin{document}

\pagestyle{plain}
\pagenumbering{roman}
\setcounter{page}{1}

\noindent
\begin{center}
{\large Roman Core Community Survey White Paper}
\end{center}

{\large {\bf Title:} Massive Black Hole Binaries as LISA Precursors in the High Latitude}

\vspace{0.1\baselineskip}

\hspace{0.52in}{\large Time Domain Survey}

\vspace{0.5\baselineskip}

{\bf Abstract:} {\it With its capacity to observe $\sim 10^{5-6}$ faint active galactic nuclei (AGN) out to redshift $z\approx 6$, Roman is poised to reveal a population of  $10^{4-6}\, \msun$ black holes during an epoch of vigorous galaxy assembly. By measuring the light curves of a subset of these AGN and looking for periodicity, Roman can identify several hundred massive black hole binaries (MBHBs) with 5-12 day orbital periods, which emit copious gravitational radiation and will inevitably merge on timescales of $10^{3-5}$ years. During the last few months of their merger, such binaries are observable with the Laser Interferometer Space Antenna (LISA), a joint ESA/NASA gravitational wave mission set to launch in the mid-2030s.  Roman can thus find LISA precursors, provide uniquely robust constraints on the LISA source population, help identify the host galaxies of LISA mergers, and unlock the potential of multi-messenger astrophysics with massive black hole binaries.}

\vspace{0.5\baselineskip}

{\bf Name of Survey:} High Latitude Time Domain Survey

{\bf Scientific Category:}  Supermassive black holes and active galaxies 

{\bf Additional scientific keywords:} Gravitational waves, Supermassive black holes, High-luminosity active galactic nuclei, Low-luminosity active galactic nuclei, Quasars

\vspace{0.5\baselineskip}

{\bf Submitting Author:}  Zolt\'an Haiman {\it (Columbia University, zoltan@astro.columbia.edu)}\\

\vspace{-0.5\baselineskip}

{\bf Contributing Authors:} 
\vspace{-0.5\baselineskip}
\author{Chengcheng Xin}
\affiliation{Columbia University, New York, USA; email: cx2204@columbia.edu}
\author{Tamara Bogdanovi\'c}
\affiliation{Georgia Tech University, Atlanta, USA; email: tamarab@gatech.edu}
\author{Pau Amaro Seoane}
\affiliation{Universitat Politècnica de València; email: amaro@upv.es}
\author{Matteo Bonetti}
\affiliation{University of Milano Bicocca, Italy; email: matteo.bonetti@unimib.it}
\author{J. Andrew Casey-Clyde}
\affiliation{University of Connecticut, Storrs, CT, USA; email: andrew.casey-clyde@uconn.edu}
\author{Maria Charisi}
\affiliation{Vanderbilt University, Nashville, USA; email: maria.charisi@vanderbilt.edu}
\author{Monica Colpi}
\affiliation{University of Milano Bicocca, Italy; email: monica.colpi@unimib.it}
\author{Jordy Davelaar}
\affiliation{Center for Computational Astrophysics, Flatiron Institute, New York, USA; email: jdavelaar@flatironinstitute.org}
\author{Alessandra De Rosa}
\affiliation{INAF, Rome, Italy; email: alessandra.derosa@inaf.it}
\author{Daniel J. D'Orazio}
\affiliation{NBIA, Copenhagen, Denmark; email: dorazio@nbi.ku.dk}
\author{Kate Futrowsky}
\affiliation{Georgia Tech University, Atlanta, USA; email: kfutrowsky@gatech.edu}

\author{Poshak Gandhi}
\affiliation{University of Southampton SO17\,1BJ, UK; email: poshak.gandhi@soton.ac.uk}

\author{Alister W.\ Graham}
\affiliation{Swinburne University of Technology, Australia; email: agraham@swin.edu.au}
\author{Jenny E. Greene}
\affiliation{Princeton University, New Jersey, USA; email: jgreene@astro.princeton.edu}
\author{Melanie Habouzit}
\affiliation{Zentrum f\"ur Astronomie, Max-Planck-Institut f\"ur Astronomie, Germany; email: habouzit@mpia.de}
\author{Daryl Haggard}
\affiliation{McGill University, Montreal, Canada; email: daryl.haggard@mcgill.ca}
\author{Kelly Holley-Bockelmann}
\affiliation{Vanderbilt University, Fisk University, Nashville, USA; email: 
k.holley@vanderbilt.edu}
\author{Xin Liu} 
\affiliation{UIUC, Urbana-Champaign, USA: email: xinliuxl@illinois.edu}
\author{Alberto Mangiagli}
\affiliation{CNRS, Paris, France; email: alberto.mangiagli@apc.in2p3.fr}
\author{Alessandra Mastrobuono-Battisti}
\affiliation{GEPI, Observatoire de Paris,  France; email:alessandra.mastobuono-battisti@obspm.fr}
\author{Sean McGee}
\affiliation{University of Birmingham; email: smcgee@star.sr.bham.ac.uk}
\author{Chiara M. F. Mingarelli}
\affiliation{Yale University, New Haven, USA; email: chiara.mingarelli@yale.edu}
\author{Rodrigo Nemmen}
\affiliation{Universidade de S\~ao Paulo, Brazil; email: rodrigo.nemmen@iag.usp.br}
\affiliation{Stanford University, USA}
\author{Antonella Palmese}
\affiliation{Carnegie Mellon University, Pittsburgh, USA; email: palmese@cmu.edu}
\author{Delphine Porquet}
\affiliation{Aix Marseille Univ., CNRS, CNES, LAM, Marseille, France; email: delphine.porquet@lam.fr}
\author{Alberto Sesana}
\affiliation{University of Milano Bicocca, Italy; email: alberto.sesana@unimib.it}
\author{Aaron Stemo}
\affiliation{Vanderbilt University, Nashville, USA; email: aaron.m.stemo@vanderbilt.edu}
\author{Alejandro Torres-Orjuela}
\affiliation{TianQin Center, SYSU, Zhuhai, China; email: atorreso@mail.sysu.edu.cn}
\author{Jonathan Zrake}
\affiliation{Clemson University, Clemson, USA; email: jzrake@clemson.edu}

\maketitle
\bibliographystyle{spphys}       

\newpage
\pagestyle{plain}
\pagenumbering{arabic}
\setcounter{page}{1}

{\bf 1. Introduction and Background}
\vspace{0.5\baselineskip}

Massive black holes (MBHs) in the $\approx 10^6-10^{10}{\rm M_\odot}$
range are present in the nuclei of most, and perhaps all, nearby
galaxies \cite{KormendyHo2013}.  In hierarchical structure formation
models, galaxies are built up by mergers between lower-mass
progenitors. Each merger event is expected to deliver the nuclear
MBHs~\cite{springel2005,2021ApJ...923..146G}, along with a significant
amount of gas~\cite{bh92}, to the central regions of the new
post-merger galaxy. The inevitable conclusion is that massive black
hole binaries (MBHBs) should form frequently in galactic nuclei over
cosmic time-scales, and that this should often take place in gas-rich
environments.  These MBHBs are both a fundamental product of galaxy
formation and evolution, and the prime targets of gravitational wave
(GW) experiments, from the space-based GW detectors LISA, TianQin, and
Taiji~\cite{LISA,tianqin,taiji}, to pulsar timing
arrays~\cite{Goulding2019, Xin2021, CaseyClyde2022, Ming2023,
  Koss2023}.  The combination of GW and EM detections of the same
population of compact MBHBs will open windows to a range of new
science, from understanding the astrophysical environments and host
galaxies of MBHBs over cosmic time, to understanding binary evolution,
accretion, and emission~\cite{MMMBA-decadal}.

The pair of MBHs from the kpc scale of a galactic collision must be
dragged down to several orders of magnitude smaller separations to
become a genuine gravitationally bound binary, and to eventually merge
due to GW emission. This requires efficient dissipation of orbital
energy and angular momentum transport by stellar and gas torques. The
time to merger can be as short as 100 Myr or as long as (or exceed)
the age of the Universe, depending on the geometry of the encounter,
the morphology of the interacting galaxies, and a variety of
inherently stochastic astrophysical processes
\cite{2023LRR-astro}. Tracking the late stages of this long journey to
coalescence, through detecting the light produced by the accretion
onto MBHBs, is key to assessing the existence of merging
systems. Hydrodynamical simulations have recently reached a consensus
that the emission from MBHBs, when surrounded by dense nuclear gas,
should be just as bright as from single MBHs, all the way to the
merger~\cite{Krauth+2023}. Additionally, binary emission is strongly
modulated on the orbital period of the binary, either due to
hydrodynamical effects, or due to relativistic Doppler
modulations~\cite{Bogdanovic+2022}.

There is, indeed, empirical evidence for pairs of MBHs in galactic
nuclei, as expected from galaxy mergers.  A few active MBH pairs
(so-called dual AGN) have been resolved at projected separations of
$0.1-1$kpc in the optical \cite{Goulding2019,2021ApJ...923..146G} and
X-ray bands \cite{Fabbiano+2011}.  A compact MBHB at a projected
separation of $\sim7$ pc, discovered in the radio \cite{rodriguez06},
is a candidate gravitationally bound system.  Some even more compact,
spatially unresolved gravitationally bound MBHB candidates have been
identified through indirect observational techniques relying on
quasi-periodicities in light-curves, spectral features such as
double-peaks or velocity offsets of emission lines, or spatial
structures of optical emission lines and of radio jets and lobes (see
\cite{DeRosa+2019,Bogdanovic+2022} for recent comprehensive reviews).
Of particular relevance to Roman, several hundred bright quasars have
been identified with significant optical periodicities
\cite{Graham+2015b,Charisi+2016,Liu+2019,Chen+2022}.  Identifying
quasars with periodic variability can nevertheless be challenging,
particularly when the MBHB orbital periods are relatively long
compared to the available observational baselines, and the incomplete
knowledge of the underlying variability of AGN can lead to false
detections \cite{Vaughan2016}.  To mitigate the risk of false
detections, it is thus crucial to be able to detect as many as
possible orbital cycles, something that Roman's High Latitude Time
Domain Survey (HLTDS) will be capable of doing for the short-period
($\sim 5-12$\,days) MBHBs that are inspiraling due to the emission of
GWs.

\newpage

The purpose of this white paper is to argue that Roman can make a
major contribution to this field by discovering a population of
compact MBHBs that are guaranteed LISA precursors.  This can be
achieved in a search, in an entirely new regime, for faint AGN with
short periods, at a combination of depth, cadence, and survey area
that is unique to Roman.  Periodicity is perhaps the most robust EM
signature distinguishing unresolved compact binaries from solitary
MBHs.  Identifying these sources will lead to unprecedented robust
constraints on the LISA population.  Roman will also allow us to
characterize the properties of host galaxies of LISA precursors prior
to LISA's launch. This will be crucial to help identify the hosts of
the future LISA MBHB mergers, as we expect LISA's sky localization
errors to be generally be too poor ($\sim 1-10 \,{\rm deg}^2$;
\cite{Mangiagli+2020}) to identify a unique host galaxy
\cite{Lops2023}.  Indeed, recent theoretical investigations indicate
that LISA binaries are preferentially hosted in disk-dominated,
gas-rich, and star-forming dwarf galaxies \cite{2023Izquierdo} that
Roman can identify.

More broadly, the combination of GW and EM detections of the same
population of compact MBHBs will open windows to a range of new
science, from understanding the astrophysical environments and host
galaxies of MBHBs over cosmic time, to understanding binary evolution,
accretion, and emission~(see a review in the Astro2020 Decadal Survey
white paper on multimessenger astrophysics~\cite{MMMBA-decadal}).

\vspace{\baselineskip}
{\bf 2. The Unique Contributions by Roman}
\vspace{0.5\baselineskip}

\begin{figure}[t]
\begin{center}
\includegraphics[angle=0,width=0.6\linewidth]{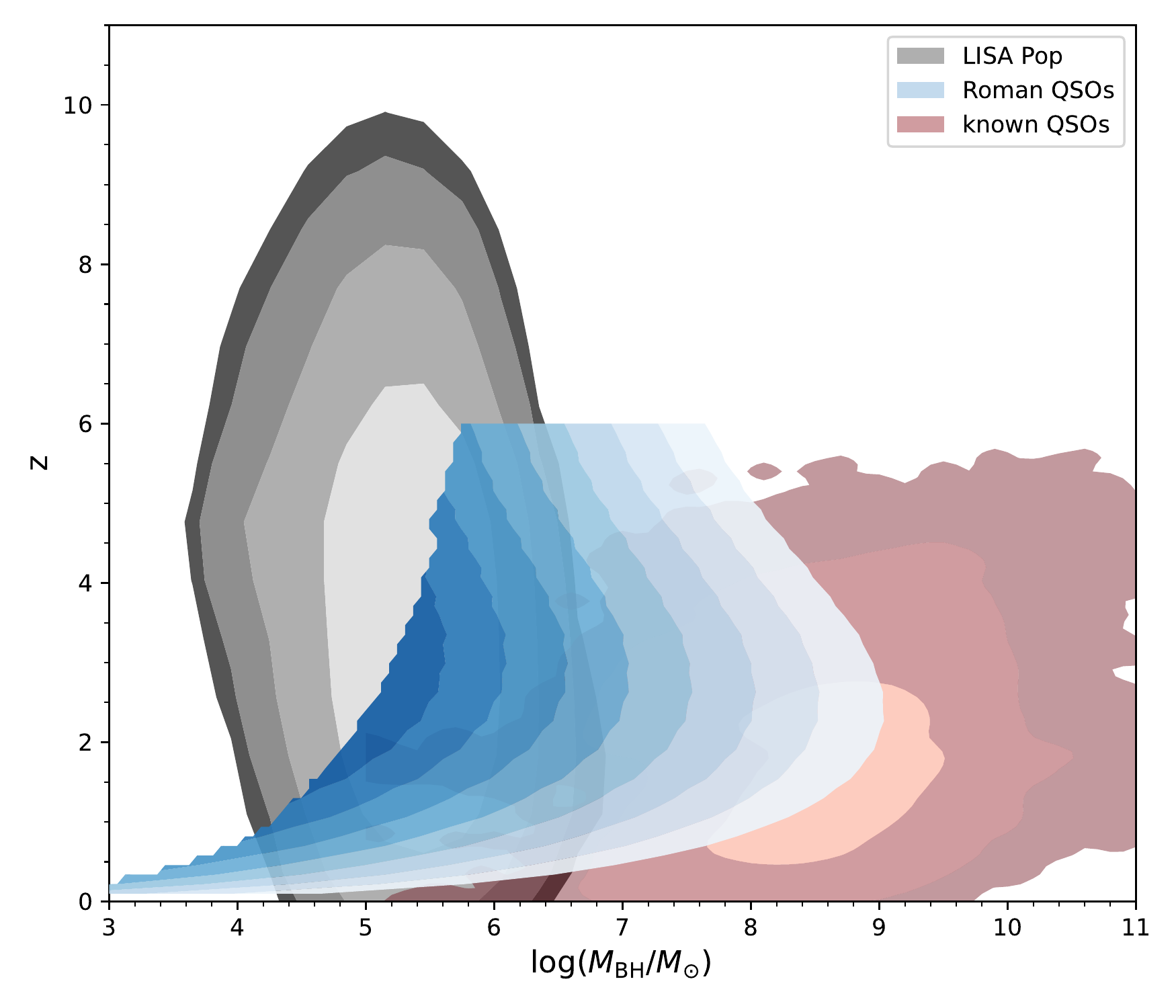}
\includegraphics[angle=0,width=0.6\linewidth]{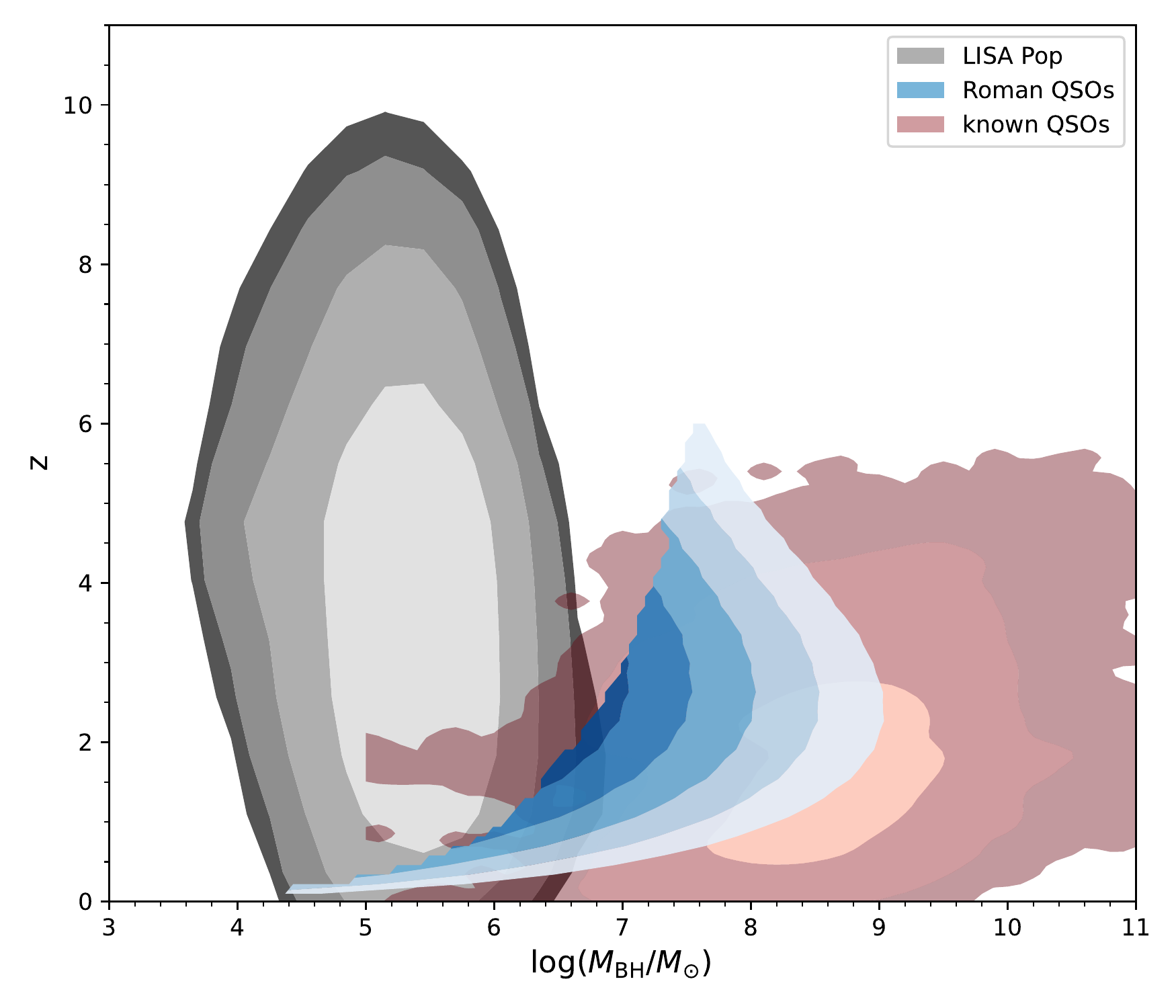}
\vspace{-1\baselineskip}
\caption{\it\small {\bf Top panel:} The distribution of massive black
  hole binary mergers expected to be detectable by LISA (a total of
  $\sim92$ sources shown by gray contours on the left; from the source
  population model in \cite{2019MNRAS.486.4044B}), together with the
  distribution of currently known quasars (contours on the right; from
  the SDSS catalog of 746,256 AGNs), as well as the AGN expected to be
  above the detection threshold by Roman, in a survey reaching a total
  co-added limit of 30 mag in the F087 band, covering 19 deg$^2$
  ($\approx 10^6$ sources in blue contours with masses above
  $10^5~\msun$; based on the extrapolation of the optical quasar
  luminosity function following \cite{Xin+2021}).  The figure
  illustrates that the current observed AGN population is too massive
  and low-redshift compared to black holes probed by LISA, but Roman
  will be able to catalog up to approximately $10^6$ black holes with
  masses and redshifts similar to those expected from LISA.  {\bf
    Bottom panel:} A subset of the faint Roman AGN are sufficiently
  bright to construct a light-curve and measure their variability.
  This panel shows, in blue contours, the mass- and redshift
  distribution of the approximately 500,000 AGN that can be detected
  in individual visits down to 28mag, along with their 10\% flux
  variations. These AGN retain significant overlap with the masses of
  the LISA sources at low redshifts.}
\label{fig:allAGN}
\end{center}
\end{figure}

Fig.~\ref{fig:allAGN} demonstrates Roman's unique synergy with LISA.
This figure shows, in the gray contours, the distribution of MBHB
mergers in total binary mass and redshift expected to be detectable by
LISA, as predicted by the semi-analytical population model of
\cite{2019MNRAS.486.4044B}.  This distribution depends on model
assumptions, such as the nature of the MBH seeds -- i.e. whether the
$\sim10^{5-6}~\msun$ MBHs grow from $10-100~\msun$ stellar-mass black
holes, or result from the rapid direct collapse of gas (the latter
case is shown in Fig.~\ref{fig:allAGN}).  Given a sufficient number of
detections, LISA will be able to probe the contributions from these
different channels.  Most importantly, the MBHs probed by LISA have
low masses, closer to the
``intermediate"~\cite{Chilingarian+2018,Graham+2019} than to the
``supermassive" range.  For reference, in Fig.~\ref{fig:allAGN} we
also show, in the purple contours, the black hole mass and
redshift-distribution of the $\approx 750,000$ AGN presently cataloged
in the Sloan Digital Sky Survey (SDSS) \cite{Yang+2021,Wu+Shen2022}.
These AGN are powered by MBHs with masses that are much higher, and
redshifts that are much lower than those of the expected LISA
population.  While there exist a small number of nearby,
low-luminosity AGN with low black hole masses comparable to those in
the LISA band~\cite{Greene+2020,2021ApJ...923..146G}, and JWST and
other instruments can also detect high-$z$ quasars overlapping with
this mass range, they are too few in number and lack sufficient
time-domain data for a systematic search for the rare short-lived
binaries among this population.  A handful of inactive MBHs in the
same low-mass range have also been detected through kinematics, but
their binarity is unknown.

In contrast, the blue contours in Fig.~\ref{fig:allAGN} show the
distribution of {\it all AGN} expected to be above the detection
threshold of Roman, when all of the data from the HLTDS are co-added.
These contours were computed by extrapolating the empirically measured
quasar luminosity function (LF), as compiled by \cite{Kulkarni2019},
to lower black hole masses and higher redshifts.  The extrapolation
methods and numerical choices are detailed in \cite{Xin+2021}.  For
the Roman survey, in the top panel, we assume a solid angle of 19
deg$^2$ and a co-added detection threshold of 30 mag in the F087
filter (this corresponds, for example, to 36 separate data points,
each with an effective depth of 28 mag), which yields a total of $\sim
10^6$ AGN.  To convert to BH mass, we assume 10\% of the Eddington
luminosity in the F087 band.  For reference, for a $10^5~\msun$ MBH at
$z=2$, the luminosity of 0.1$L_{\rm Edd}$ corresponds to 30 mag.

As Fig.~\ref{fig:allAGN} illustrates, a sufficiently deep Roman survey
is sensitive to the same black hole masses as LISA (albeit somewhat
lower redshift). We note that this capability is unique to Roman's
depth -- for example, the Legacy Survey of Space and Time (LSST) by
the Vera C. Rubin Observatory will have a $\sim 4$ magnitude shallower
co-added depth ($\sim 26$ mag in the $i$ band, closest to Roman's
F087), and will accordingly probe nearly $\sim$two orders of magnitude
higher black hole masses~\cite{Ivezic2019}.  As a result, Roman's
HLTDS has the unique capability to discover essentially the same
population of MBH binaries as LISA, but at somewhat earlier stages of
their merger, with larger separations.

However, depth is not the only necessary ingredient to accomplish
this.  First, in order to search for periodicities, the effective
depth is reduced for two reasons: (i) the AGN needs to be detected at
each visit, not just in the full survey co-add, and (ii) its
variability needs to be measured (for example, measuring 10\% flux
variations requires $\approx 2.5$mag deeper photometry than a mere
detection).  Imposing these more stringent effective photometric
limits leaves us with $\approx 550,000$ AGN, shown in the bottom panel
of Fig.~1. This bottom panel adopts a sensitivity threshold of 28 mag
in the F087 filter, and assumes that 10\% of the band flux is
variable, effectively assuming a luminosity of $L=0.01L_{\rm Edd}$ in
the F087 band.

In order to search for the compact MBHB LISA precursors, identified by
their periodical variability among these AGN, solid angle and cadence
are also critical. To be a guaranteed LISA precursor, an MBHB needs to
be sufficiently compact to be safely in the GW-driven regime, where
the inspiral time is at most $10^{4-5}$ years, and where the binary
cannot be disrupted (although circumbinary gas disk torques could
still be a significant contribution and affect the inspiral
time;~\cite{Haiman+2009}). As a result, these short-lived sources will
constitute a small fraction of all AGN (requiring a minimum solid
angle), and will have short periods (also requiring a minimum
cadence).

To illustrate these points, we note that the $\approx$ 300 MBHB
candidates, identified among large time-domain surveys of bright
quasars based on their
periodicity~\cite{Graham+2015b,Charisi+2016,Liu+2019,Chen+2022},
remain controversial. This is primarily because all AGN are known to
be variable, and stochastic red noise (the so-called ``damped random
walk''; \cite{Macleod2010}) can mimic periodicity when only a few
cycles are observed.  The existing large time-domain surveys are
shallow, and the masses of the periodic MBH candidates are
$\approx10^9~\msun$.  These heavy binaries inspiral very rapidly.  The
GW inspiral time scales with chirp mass (the combination $M_{\rm
  chirp}\equiv(M_1M_2)^{3/5}/(M_1+M_2)^{1/5}$ of the masses of the
individual MBHs) as $t_{\rm GW}\propto M_{\rm chirp}^{-5/3}$, and the
scaling with an orbital period is even steeper; $t_{\rm GW}\propto
P_{\rm orb}^{8/3}$.  For an equal-mass binary with
$\approx10^9~\msun$, source redshift $z\approx 2$, and observed
orbital period of $P_{\rm orb}=1-5$ years, the GW inspiral times are
$\sim 10^{3-5}$ years. As a result, shorter-period binaries are too
rare to be found in these surveys.  Because of this, the existing MBH
binary candidates only cover at most a few cycles and remain possible
to attribute to red noise, as mentioned in the Introduction.

The combination of its depth, solid angle, cadence, and the total
duration will allow Roman's HLTDS to extend the search for periodic
AGN to an entirely new regime. Taking 10$^5~\msun$ and $z=1$ as the
approximate black hole mass and redshift corresponding to a typical
Roman AGN that can be identified as periodic (see Figs.~2 and 3
below), MBHBs with this mass inspiral in $\approx$10$^{3-5}$ years if
their observed orbital periods are $P_{\rm orb}=5-12$ days.  Such
short periods will allow the detection of a large number of cycles and
securely rule out fake periodicity caused by red noise.  In
particular, red noise cannot mimic periodicity when more than $\approx
10$ cycles are measured~\cite{Vaughan2016}, requiring a total of 4
months of observation time for a 12-day observed period.  We note
that, depending on mass ratio and circumbinary disk properties, the
dominant optical/IR periodicity from hydrodynamical modulations by a
binary could appear on timescales a few times longer than the orbital
period~\cite{Dorazio+2015}. Roman's light curves will need to be
carefully analyzed for the presence of multiple periods -- this would
help probe the binary system's parameters if more than one period is
detected but would require longer baselines if only the longer periods
are visible in the data.

In the GW-driven regime, discovering several dozen or more sources
with a range of periods will allow a novel test of the GW inspiral
time, since the number of sources should scale with their period as
P$^{8/3}$ -- while deviations from this scaling could be used to study
the torques from circumbinary disks~\cite{Haiman+2009}. Most
importantly, identifying this short-period MBHB population will
provide a unique synergy between Roman and LISA: these sources are
guaranteed precursors to LISA events, and will yield robust,
model-independent constraints on the LISA merger rate.

Finally, how many periodic AGN might Roman's HLTDS be able to
identify?  One approach is to model the merging MBH population,
building on hierarchical galaxy formation models, calibrated to
observations, such as the empirically determined black hole-galaxy
mass scaling relations.  A sequence of parallel quadratic scaling
relations, which track the evolution of galaxies toward increasingly
bulge-dominated morphologies, has been established and can be a key
ingredient in such modeling~\cite{2023MNRAS.522.3588G}.  A simple
quick way to obtain an upper limit on this number, without such
modeling, is to assume that AGN are often activated by mergers, so
that a fraction $f_{\rm bin}\sim1$ of all AGN correspond to MBHB
mergers.  In this case, the number $N_{\rm bin}$ of binaries with a
particular orbital period, and corresponding inspiral time $t_{\rm
  GW}$, is simply given by the fraction of AGN `caught' at this stage:
$N_{\rm bin}=[t_{\rm GW}$ / $t_{\rm Q}] f_{\rm bin} N_{\rm AGN}$,
where $t_{\rm Q}=10^{7-8}$ years is the typical quasar
lifetime~\cite{Martini2004} and $N_{\rm AGN}$ is the total number of
AGN in a survey.  In our fiducial case, Roman's survey will contain
$N_{\rm AGN}\approx10^5$ AGN whose periodicities could be measured, so
that inspiral times of $10^{3-5}$ years will yield $(10-1000) f_{\rm
  bin}$ binaries with observed periods of 5-12 days for $t_{\rm
  Q}=10^{7}$ years, and $(1-100) f_{\rm bin}$ binaries if $t_{\rm
  Q}=10^{8}$ years.  This simple approach yields approximately the
correct fraction $\sim10^{-3}$ of binary candidates among the bright
quasars that have been searched to date ($\sim 300$ candidates among
300,000 quasars). On the other hand, it likely represents an upper
limit, because if all present candidates were real, they would, in the
simplest models, overpredict the stochastic GW background at nHz
frequencies~\cite{Sesana+2018}.

The results of the actual calculations of these numbers, i.e. using
the quasar LF, assuming a quasar lifetime of $t_{\rm Q}=10^7$ years,
and including the GW inspiral time at a fixed observed period as a
function of binary mass and redshift, are shown in Figs.~2~and~3.
Fig.~2 shows the distribution of Roman AGN with an orbital period of 6
days (top) or 12 days (bottom), assuming a single-visit depth of 28
mag in a 19 deg$^2$ survey.  As the figure shows, these periodic
sources preferentially sample lower masses and redshifts among the
Roman AGN population.  These biases, compared to the parent population
of all Roman-detected AGN, are caused by the strong dependence of the
GW inspiral time on chirp mass and orbital time. In particular, at a
fixed observed period, more massive binaries, as well as binaries at
higher redshift (and therefore with shorter rest-frame periods)
inspiral more rapidly and are correspondingly rarer.  Conservatively
excluding black holes with masses below $10^5~\msun$ (since the
existence and abundance of these intermediate-mass systems is less
well understood), we find a total of 348 ($P_{\rm orb}$=6 days) and
2012 ($P_{\rm orb}$=12 days) periodic sources.  Fig.~3 shows the same
result, except for a shallower depth of 27 mag, reducing the numbers
by a factor of 3-4, to 102 ($P_{\rm orb}$=6 days) and 531 ($P_{\rm
  orb}$=12 days) sources.

\vspace{\baselineskip}
{\bf 3. Survey Parameter Considerations}
\vspace{0.5\baselineskip}
 
The fiducial survey, discussed in the previous section, assumes a
depth of 28 mag in the F087 filter (for each single data point in the
light-curve), a solid angle of 19 deg$^2$, a total observation time of
at least 4 months, and sufficient cadence to identify 5-12 day
periods, i.e. of order 5 days.  At least three different filters would
be ideal in order to compare the light-curves in different bands, and
help distinguish periodic binary variability from other possible
sources of AGN periodicity.  While variability itself will be the most
efficient way to select AGN to begin with~\cite{IvezicMacLeod2014},
having colors may also help this selection, especially if, using
external space-based data, the host galaxies can be resolved and
subtracted.

In general, the considerations driving survey specifications for MBHB science are as follows:

\begin{itemize}

\item {\bf Depth.}  This is a primary consideration: the depth needs
  to be sufficient to detect faint AGN, corresponding to black hole
  masses in the LISA regime.  Since we require identifying
  periodicities (not just detecting an AGN), this depth requirement is
  for each individual visit, and needs to be at least $27-28$ mag in
  the F087 filter. This is the closest among Roman's filters to the
  LSST $i$ band considered in the analysis in \cite{Xin+2021}. The
  depth requirement is set by assuming black holes emit a luminosity
  of $L=0.1 L_{\rm Edd}$ in the F087 band, of which 10\% is variable
  (effectively requiring a depth corresponding to $L=0.01 L_{\rm
    Edd}$).

\item {\bf Solid Angle.}  The choice of solid angle is driven by
  ensuring that we have a sufficiently large sample of AGN among which
  the rare compact binaries, with inspiral times of $10^{3-5}$ years,
  will be represented.  This necessitates a total sample size of at
  least $10^5$ AGN, requiring a solid angle of $\gsim 5$ deg$^2$.  The
  number of objects scales simply linearly with survey size.

\item {\bf Cadence.}  The choice of cadence is driven by ensuring that
  periodicities as short as 5 days can be inferred from the time
  series.  Naively, this requires a cadence of 5 days or more
  frequent. In principle, however, the Nyquist frequency can be beaten
  if the sampling is irregular, therefore, more sparse sampling is
  tolerable, as long as the observations are not placed at regular
  intervals.  Finally, a single (or at most a few) long contiguous
  chunks of data are preferred over many separate campaigns spread
  over a longer baseline. Long gaps make the modeling of the red noise
  harder, which will increase the false detection rate.

\item {\bf Total duration.}  The total duration is driven by requiring
  the detection of at least 10 cycles -- for the 12-day period
  binaries, this requires at least 4 months of total coverage. Longer
  durations would improve sensitivity to longer periods and would
  significantly increase the total sample of periodic AGN (since the
  number increases as steeply as $P_{\rm orb}^{8/3}$ for binaries in
  the purely GW-driven regime).

\newpage
\item {\bf Sky Location.}  In principle, any location on the sky is
  acceptable.  However, it would be useful to choose areas already
  covered by the deepest existing optical and infrared fields, such as
  COSMOS or Subaru's SHELLQs surveys, which already contain data on
  faint AGN.  In particular, with Roman's restriction of being within
  36 deg of the poles, there are several deep multi-wavelength
  observations near the South Galactic Pole.  Of specific interest is
  the deep multi-wavelength GOODS-S field, which includes Hubble Deep
  and Ultradeep, Chandra Deep, Spitzer Deep, and will include LSST
  Wide and Deep, as well as Euclid Deep observations. The CANDELS team
  has published recent multi-wavelength catalogs for
  GOODS-S~\cite{Barro+2019,Kodra+2023} and specific deep AGN catalogs
  may be available in the near future. The locations of the LSST deep
  drilling fields would be other good possibilities.

\item {\bf Filter choices.}  The filter choices are primarily driven
  by the need to distinguish periodic fluctuations caused by MBHBs
  from other sources of AGN periodicities.  The low-mass AGN are
  expected to be bright in the F087 filter, and two additional bands
  would allow the construction of color-color diagrams.  Measuring the
  chromaticity of the periodic variability can then be compared to
  theoretical predictions for binaries~\cite{Westernacher+2022}. The
  more the filters are spread over wavelength (to obtain larger 'lever
  arms' in color-color space), the better.  Color-color diagrams could
  possibly help with AGN selection but would likely require external
  space-based data to resolve and subtract the host galaxy.

\end{itemize}

\vspace{\baselineskip}
{\bf 4. Summary and Conclusions}
\vspace{0.5\baselineskip}

The fiducial baseline HLTDS, planned around the requirement to search
for Type Ia SNe, consists of 6 months of observing time, spread over 2
years, with a cadence of 5 days, in three filters, covering 5 deg$^2$,
to a depth of $\sim$26 mag.

This turns out very similar to the requirements outlined here for a
search for periodic, faint AGN containing MBHBs that are LISA
precursors.  The specification where we most significantly push on the
above baseline survey is depth: single-visit exposures need to reach
27-28 magnitudes in order to detect faint AGN with black hole masses
in the LISA band ($10^{5-6}~\msun$).  On the other hand, a somewhat
sparser cadence can be tolerated, as long as the sampling is irregular
in time (to beat the Nyquist frequency).

With the detection of a small population of LISA precursors, Roman can
make a major contribution to GW and multi-messenger science with
massive black hole binaries. There are additional ways in which Roman
can contribute to GW science, which we list in the Appendix for
completeness.

\clearpage
\newpage

\clearpage
\newpage
\begin{figure}[t]
\begin{center}
\includegraphics[angle=0,width=0.6\linewidth]{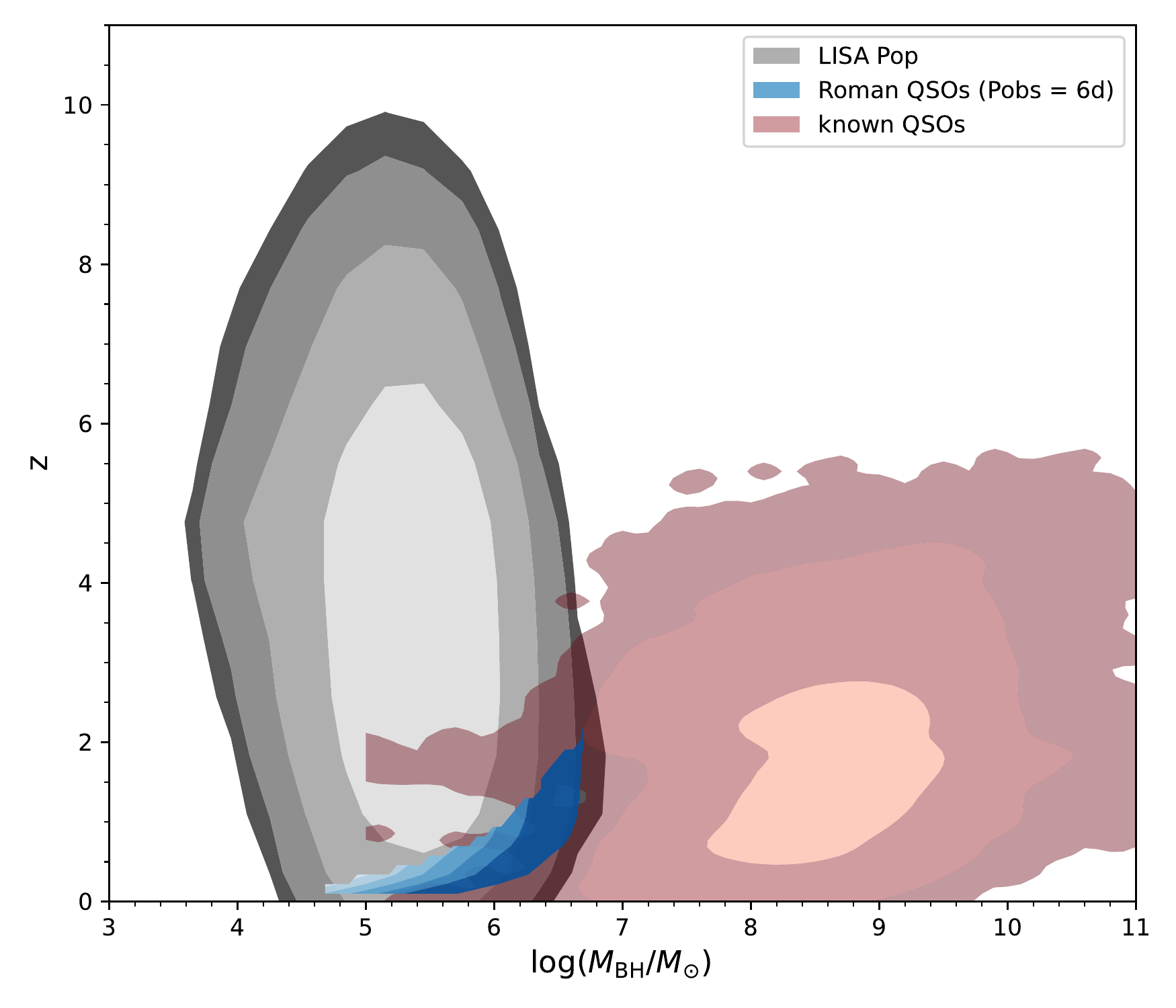}
\includegraphics[angle=0,width=0.6\linewidth]{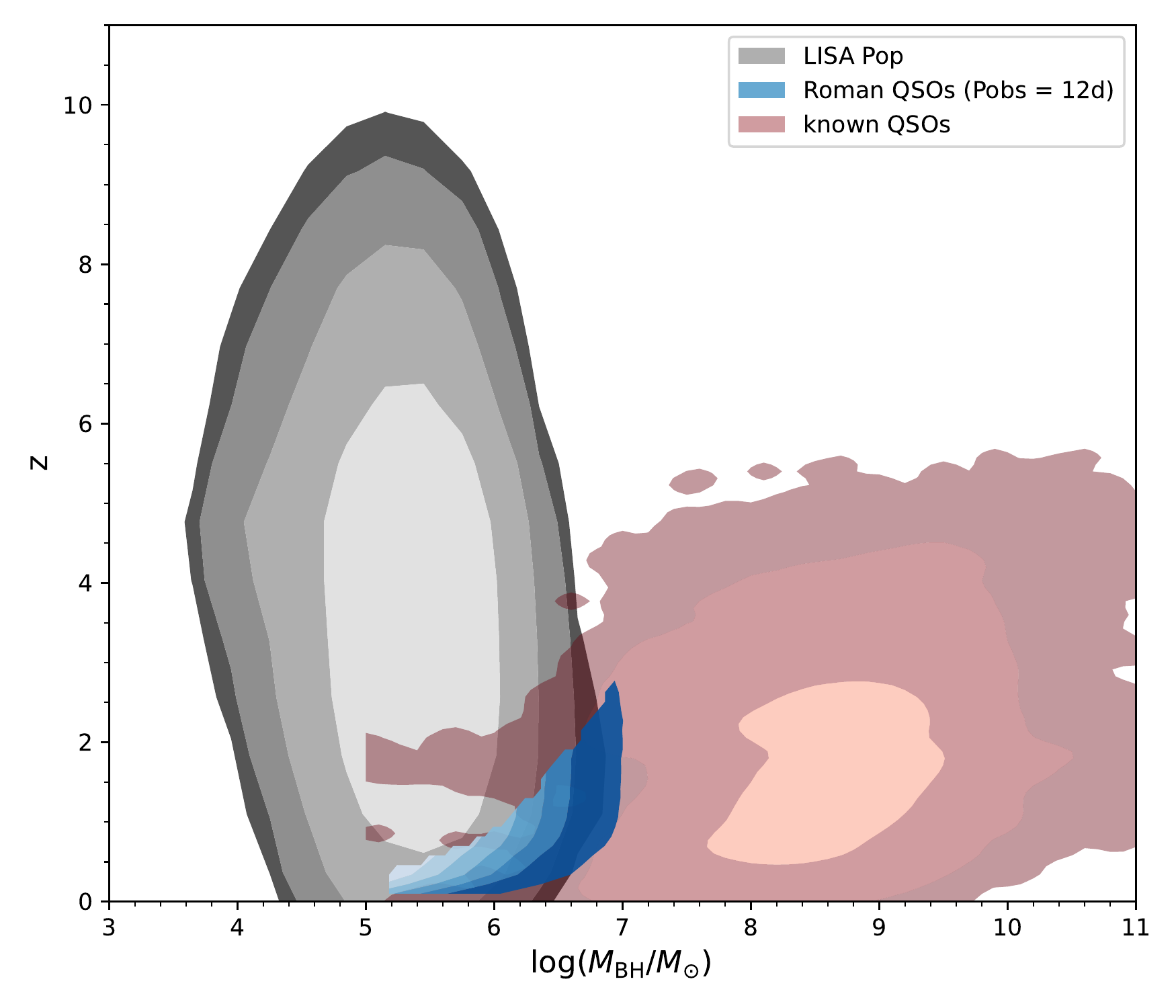}
\vspace{1\baselineskip}
\caption{\it\small The distribution of Roman AGN with an orbital period of 6 days (top) or 12 days (bottom), assuming single-visit depth of 28 mag in a 19 deg$^2$ survey.  The GW inspiral time of these sources are $\sim 10^4$ yrs and $\sim 10^5$ yrs respectively.  At these short inspiral times, they represent a small fraction of the total Roman AGN population from Fig.1, yielding 348 ($P$=6 days) and 2012 ($P$=12 days) sources that are {\bf guaranteed} LISA precursors.}
\label{fig:periodicAGN}
\end{center}
\end{figure}

\clearpage
\newpage
\begin{figure}[t]
\begin{center}
\includegraphics[angle=0,width=0.6\linewidth]{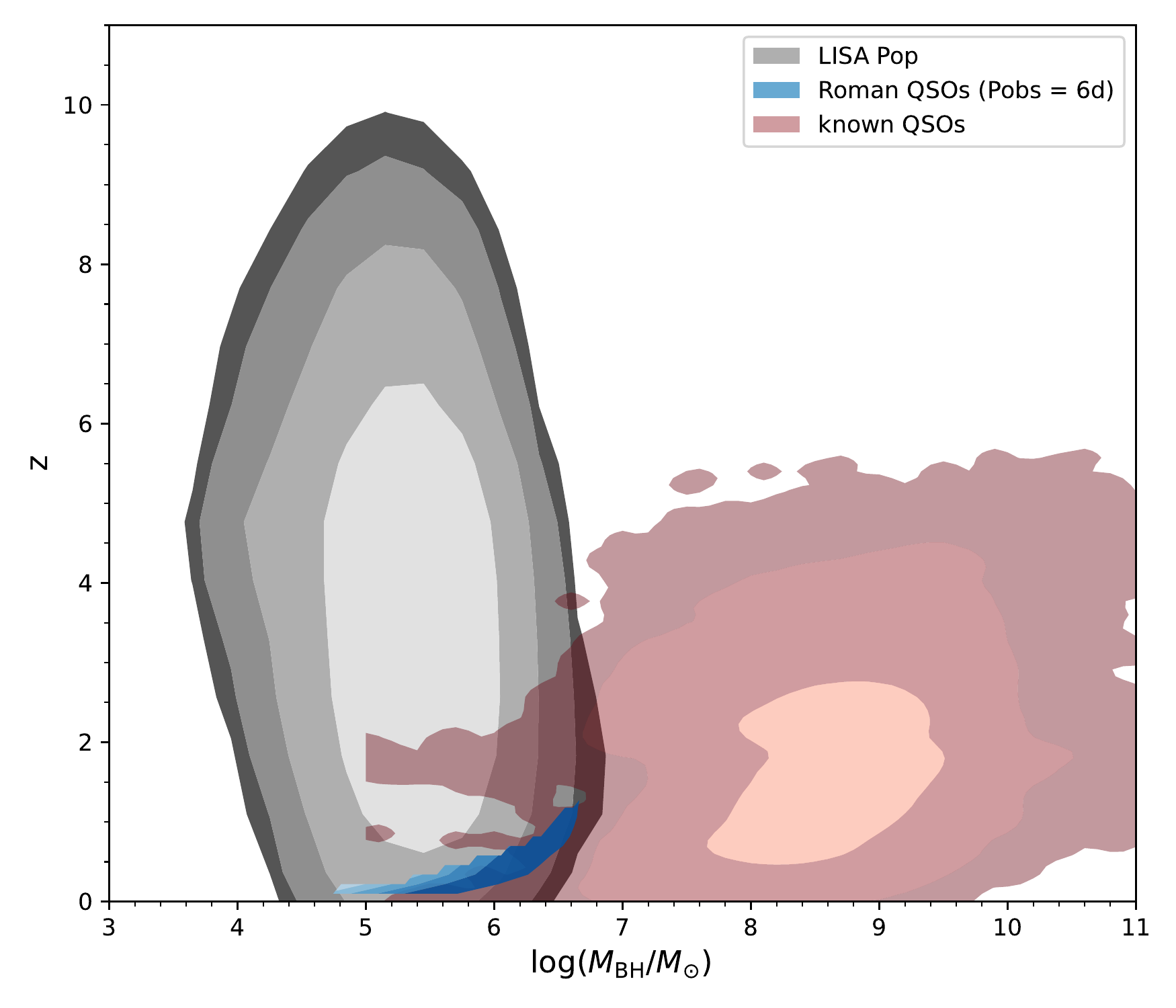}
\includegraphics[angle=0,width=0.6\linewidth]{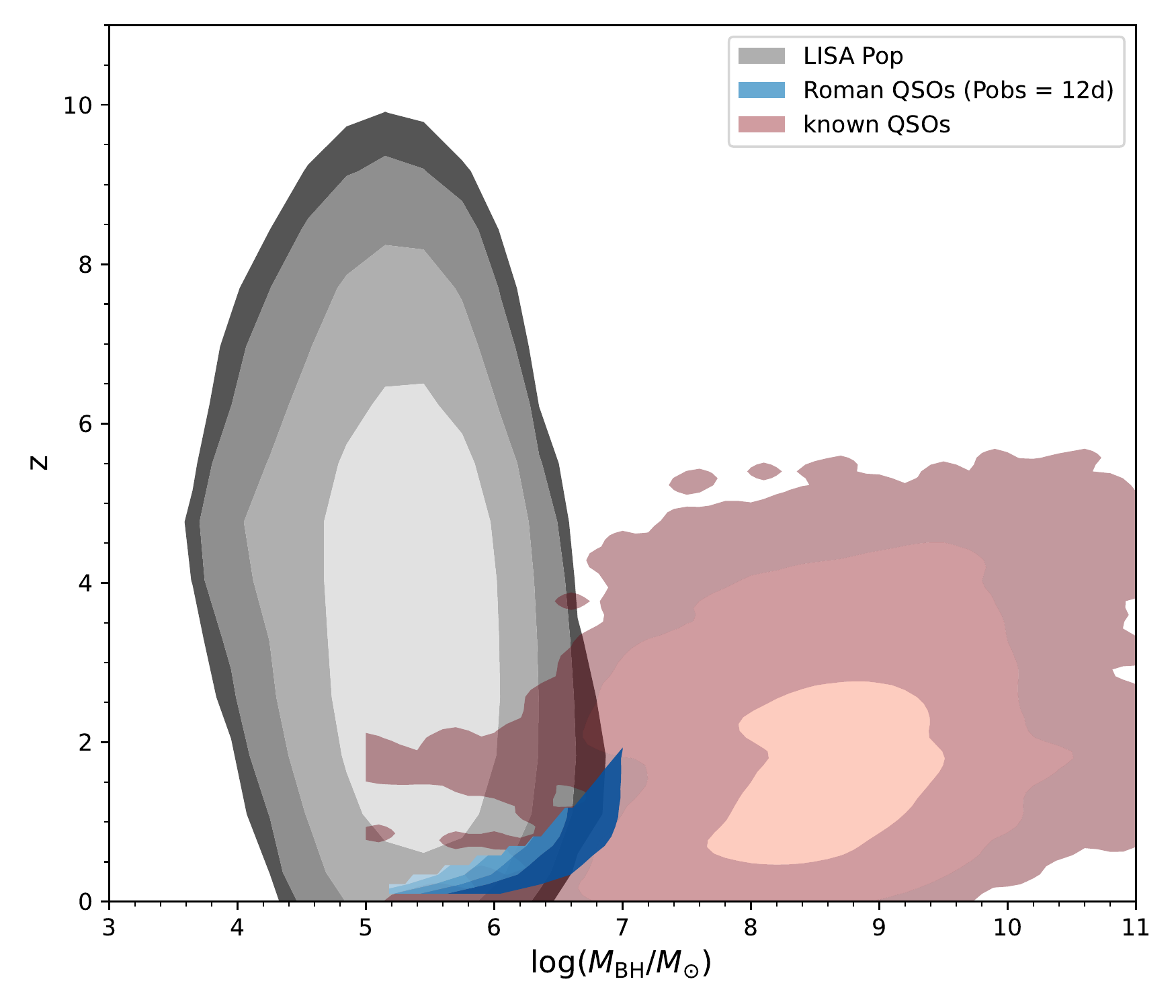}
\vspace{1\baselineskip}
\caption{\it\small Same as Fig.~\ref{fig:periodicAGN} but assuming a
  shallower 27-mag single-visit sensitivity; this reduces the
  detectable sample of LISA precursors by a factor of 3-4 to 102
  ($P$=6 days) and 531 ($P$=12 days).}
\label{fig:periodicAGN-27mag}
\end{center}
\end{figure}

\clearpage
\newpage
{\bf Appendix:  Additional GW Science with Roman}
\vspace{0.5\baselineskip}

The survey design above should allow the identification of LISA
pre-cursors.  There are other ways in which Roman can contribute to GW
science with a survey similar to the one proposed here, as well as
with its other Core Community Surveys and even beyond.  Here we list
these additional possible science investigations, and note some of the
modifications they would require to the baseline survey above.

\vspace{\baselineskip}

{\it (i) LIGO sources in AGN disks.} From the observation of
stellar-mass binary black holes (BBHs) by LIGO and Virgo, we can infer
a merger rate of $10 - 100\,{\rm Gpc^{-3}\,yr^{-1}}$~\cite{LIGO_2020},
where it is estimated that 25\,\% to 80\,\% of these binaries form in
AGN disks~\cite{Ford_McKernan_2022}. LIGO will see BBHs out to a
distance of $2.5\,{\rm Gpc}$ from its fifth observing run (O5;
starting around 2025)~\cite{LIGO_2020}.  Assuming that Roman's HLTDS
will use six months of aggregate telescope time, we anticipate around
$80 - 2,620$ BBHs in AGN disks to merge during this period. After the
BBH merges, the remnant black hole gets gravitationally
kicked~\cite{pretorius_2005} and its motion through the gas of the AGN
disk triggers a bright flare that is briefly comparable to or can
exceed the AGN background in the near-infrared and
optical~\cite{rodriguez-ramirez_et_al_2023,Tagawa+2023}. The flare
becomes detectable around $20-500$ days after the GW event and lasts
for days to months. Candidates for such events have been identified by
LIGO/Virgo and the Zwicky Transient Facility, including GW190521 and
ZTF19abanrhr~\cite{graham_et_al_2020}, as well as a handful of
additional coincident flares~\cite{Graham+2023}.  We anticipate that
in the best case, tens of BBHs residing in AGN disks will merge in the
area observed by the Roman Space Telescope but in the worst case, no
event might be detected. If Roman extends the duration of the HLTDS --
in particular, after 2026 when LIGO, Virgo, and KAGRA should further
improve their detection distance -- it might be able to detect
hundreds of these AGN flares.

\textbf{\it (ii) Self-lensed binaries.}  In addition to
Doppler-induced modulations in the lightcurves of accreting MBHBs,
there is a general relativistic effect that could also be detectable
by Roman. If a massive binary is observed close to edge-on, the
foreground BH's Einstein ring eclipses the background BH, resulting in
a gravitational lensing (or self-lensing flare; SLF) event which can
enhance the observed emission substantially and serve as strong
evidence for an MBHB, see \cite{DOrazioDiStefano2018, HuDOrazio2020,
  Ingram2021, DavelaarHaiman2022}. An extensive study by
\cite{Kelley2021} showed that up to a hundred SLFs could be detected
by LSST, which raises the question if Roman could find additional SLFs
on shorter time scales.

Given a binary mass range of $10^5-10^6~\msun$, a mass ratio range of
$q=0.1-1$, an orbital period range of $P_{\rm orb}=6-12$~days, and
assuming that the light from the secondary is lensed by the primary,
an SLF, with $\geq 34\%$ magnification, has a $1-5\%$ probability for
orbital alignment and a corresponding lensing timescale of
approximately $0.5\%$ to $2.5\%$ of the orbital period
(hours). Considering finite source-size effects, maximum
magnifications may reach from $\sim10\times$ to $250\times$, though at
a lower probability for such optimal alignment
\citep{DOrazioDiStefano2020}.  So if $\gtrsim 100$ binaries are
observed, then at least one should be aligned favorably for strong
self-lensing.

However, the cadence may need to be increased from the fiducial value
in order to capture these short-duration SLFs. For binary mass $10^6
\msun$, $P=6$~days, and $q=0.1$, the duration of a SLF is $\sim0.14$
days. The probability of capturing a flare is approximately the ratio
of SLF time to orbital time, times the number of samples per orbit,
times the number of orbits sampled. Given two samples per orbit (3-day
cadence), at random phases, and 20 cycles in 4 months, one is left
with a $95\%$ chance of sampling the SLF. In this case, the
probability for favorable alignment is $\sim5\%$, so for 100 binaries
in the sample, there could be multiple lensing systems increasing the
likelihood of sampling the flare in one of them. If a periodic signal
is identified with Roman, then targeted, high-cadence follow-up could
test for the existence of flares since their relative phase in the
light curve is predicted by the model.

\textbf{\it (iii) Partial tidal disruption events.}
The survey envisioned above would also enable the discovery of
periodic flares from partial tidal disruption events (TDEs;
\cite{Payne2021}), which would be relevant for understanding the
population statistics of extreme mass-ratio inspirals (EMRIs), another
prime class of targets for LISA.

\textbf{\it (iv) Synergy with pulsar timing arrays.}  Pulsar Timing
Array (PTA) experiments are expected to detect a gravitational-wave
background sourced from the cosmic merger history of supermassive
black hole binaries with $M\gsim 10^8~\msun$. This detection can begin
to constrain the abundance of MBHs at the high-mass end. Combined with
the determination of the number density lower-mass MBHs in the Roman
survey, this will yield a constraint on the slope of the
(super)massive black hole mass function over the whole range of
$10^6-10^9~\msun$.

\textbf{\it (v) Galactic-scale dual AGN.}  Roman's High Latitude Wide
Area Survey, with its unprecedented combination of sensitivity,
spatial resolution, area, and NIR wavelength coverage, will
revolutionize the study of galactic-scale dual AGN.  The science
opportunities and technical requirements on the discovery and
characterization of SMBH pairs down to sub-kpc scales are summarized
in a separate white paper~\cite{Roman-WP-Shen}.

\textbf{\it (vi) Astrometric GW detection.} Roman may be able to
partner with LISA in yet another important way: by acting as a GW
detector of individual MBHBs and their stochastic GW background, in
the complementary microhertz range
\cite{2021PhRvD.103h4007W,2022PhRvD.106h4006W}. The Galactic Bulge
Time Domain (GBTD) survey will be particularly suitable for this
purpose given its high cadence of observations (15 minutes) and high
observed stellar density, coupled with Roman's high relative
astrometric precision. The GBTD survey requirements necessary for the
detection of GWs by Roman are described in a separate white
paper~\cite{Roman-WP-Pardo}.

\textbf{\it (vii) Simultaneous multimessenger detection of MBHBs by
  Roman and LISA.}  It is worth noting that Roman may be able to make
a unique contribution to the detection of inspiraling and merging
MBHBs beyond the bounds of its Core Community Surveys.  However, there
is an impact on Core Community Surveys, in that this would require
flexibility to interrupt ongoing surveys for target-of-opportunity
observations. Specifically, if Roman achieves a lifetime of 10 years
or longer, it is likely to operate contemporaneously with the LISA GW
observatory. In this scenario, Roman's Wide Field Instrument (WFI)
could be used for direct electromagnetic follow-up of the GW sources
detected by LISA, in cases where LISA's localization uncertainties are
not much larger than the WFI field of view. For example, we estimate
that about 50\% of MBHBs with total mass below $10^6\,M_\odot$
detected by LISA at $z=0.1$ will have their position on the sky
determined with sufficient precision to also be in the field of view
of WFI 1\,day before they merge. For MBHBs out to higher redshifts
($z\leq 1$), the WFI will fully enclose the LISA uncertainty region on
the sky for MBHBs with mass less than $10^7\,\msun$ at the time close
to the merger (i.e., within an hour from merger). In many of these
cases, the number of galaxies enclosed within the WFI field of view
will be $\sim 1-100$, providing a high likelihood for identification
of the MBHB host galaxy. The opportunity to characterize MBHB host
galaxies would provide invaluable information about their properties
and would be a unique opportunity for Roman to play a major role in
multimessenger discoveries. For these types of detections, Roman's
ability to follow up LISA sources will depend on its ability to
process LISA alerts in real-time and react to updates that provide
improved localization of MBHBs on the sky (i.e., localization that
will land within the WFI field of view). They will also depend on
Roman's agility: the ability to interrupt any other ongoing
observations and slew to the location of interest in time to catch the
two MBHs in the final act of merging.

\vspace{\baselineskip}

\clearpage
\newpage
\bibliography{refs}

\end{document}